\documentstyle[pmart,twoside,fleqn,epsf]{article} 

\begin{document}            

%
%
\title{Scattering theory of the Johnson spin transistor} 

\author{Linda S. Geux, Arne Brataas, and Gerrit E.W. Bauer}

%
%
\address{Delft University of Technology, Laboratory of Applied Physics and DIMES,\\
2628 CJ Delft, The Netherlands}
%
%

\date{\today}

\maketitle                   
%
%

\begin{abstract}
  We discuss a simple, semiclassical scattering theory for spin-dependent transport in a
many-terminal  formulation, with special attention to the four terminal device
of Johnson referred to as spin transistor. 
\end{abstract}

%
%
\section{Introduction}

Transport in ferromagnets and ferromagnetic multilayers has has become a popular subject of research since the
discovery of the giant magnetoresistance (GMR) effect,
but has been studied long before. Tedrow and
Meservey \cite{citmeservey} showed that the tunneling current in 
superconductor/insulator/ferromagnet junctions
is spin dependent. Under an applied bias, the current of spin-up and spin-down  electrons injected from a
ferromagnet into a normal metal is different. This causes a non-equilibrium magnetization
or ''spin  accumulation'' which gives rise
to an additional  boundary resistance, the ''spin-coupled interface
resistance''~\cite{citjohnsils,citson}. Johnson and
Silsbee~\cite{citjohnsils} detected this spin accumulation by two ferromagnetic
contacts to a normal metal. Spin injection can be the physical basis of new
devices, like the so-called spin transistor~\cite{citjohn,citjohnscience}. 
A "pedagogical model" of this device is shown in Fig.~\ref{spininj}(a). It
consists of a normal metal film sandwiched between two ferromagnets. The
magnetizations of the  two ferromagnets are either aligned  parallel or antiparallel. A
spin-polarized current is injected from the  first ferromagnet F1 to a drain contact P
connected to the normal metal. The  second ferromagnet F2 is attached to a floating voltage
probe. The polarized  current creates a non-equilibrium magnetization in the normal metal,
which  splits the chemical potentials of the spin-up and spin-down electrons. Just  as F1
can be seen as a spin polarizer, F2 can be seen as a spin detector, at which the chemical
potential aligns with that of the spin-up (spin-down)  electrons in the normal metal in the
parallel (antiparallel) configuration.  The voltage measured at F2 is used to determine the
impedance difference  between the parallel and the antiparallel configuration, which is
related to  the spin-coupled resistance. The experiments are actually carried out with an additional
normal metal counter electrode as in Fig.~\ref{spininj}(b). Spin-dependent transport is affected
by the properties of the normal metal.  Spin-flip scattering in the normal metal mixes the two
spin channels and  thus decreases the spin accumulation. The spin-orbit interaction
\cite{cityafet1} appears to limit the performance of the spin transistor. We discuss in this
paper the multi-terminal Landauer-B\"uttiker scattering theory of transport for a magnetic
system. This formalism provides an alternative view on the physics of the device
\cite{citdatta,citgijs} which in the diffusive regime is equivalent to previous two-terminal theories by
Johnson and Fert and Lee
\cite{citfertlee}, but we demonstrate that it can be extended to other regimes as well. For
technical details we refer to
\cite{linda}.

\begin{figure}
\label{spininj}
\vspace*{7 cm}
\caption{Schematic picture of the spin transistor for
(a) the pedagogical model and (b) the actual configuration.
}
\end{figure}

\section{The spin transistor in the Landauer-B\"uttiker formalism}

In a two-terminal configuration the current from a contact A to a contact B is 
$ I=G(\mu _A-\mu _B)e$, 
where $\Delta \mu =\mu _A-\mu _B$ is the difference in chemical potential
between the two contacts. The linear response conductance
$G$ is related to the transmission probability by the Landauer formula \cite
{citdatta}: 
\begin{equation}
\label{eqlanbut}G=\frac{e^2}h \sum_{nm,\sigma \sigma '} |t_{m\sigma , n \sigma '
}|^2, 
\end{equation}
where $t_{m\sigma , n \sigma '}$ is the transmission amplitude of an electron  
from mode $m$ with spin $\sigma$ in lead A
to mode $n$ and spin $\sigma '$ in lead B, all states being at the Fermi energy.
Here we assume for simplicity that the
magnetizations are collinear, i.e. parallel or antiparallel to each other.
A network theory of magnetic multi-terminal devices with general magnetization
is presented in \cite{network}.

B\"uttiker extended the two-terminal Landauer
formula to a many terminal device by summing over all contacts 
\cite{citdatta}. If $T_{A\rightarrow B}$ is the total transmission probability from contact
A to B: 
\begin{equation}
T_{A\rightarrow B}= \sum_{\sigma \sigma '}
\sum _{n}^{N_B}\sum_{m}^{N_A}|t_{m\sigma , n \sigma '}|^2, 
\end{equation}
where $N_A$ is the number of modes in contact A and $N_B$ the number of
modes in contact B. The total current in contact A is given by 
\begin{equation}
\label{eqmulticontact}
I_A=\frac {e^2} h\left[ \sum_{B(B\neq A)}
T_{A\rightarrow B}\mu _A- \sum_{B(B\neq A)}T_{B\rightarrow A}\mu
_B\right] . 
\end{equation}

Referring to Fig.~\ref{spininj}(a), we call the
transmission probability from F1 to F2 $T_1$. Those from the ferromagnetic contacts F1 and F2
to the drain contact of the paramagnetic film are taken to be the same $T_2$. When the net current into F2 is
zero, the potential of F2 can be obtained using Kirchhoff's law:
\begin{equation}
V_s=V\frac{T_1}{T_1+T_2}. 
\end{equation}
$V_s$ depends on the choice of
the potential zero, here taken to be here in the drain contact reservoir. On the other hand,
$Z_s$, the impedance between F1 and P, only depends on the potential difference $V$.
A constant current $I$ is driven from F1 into the drain contact of the
paramagnetic film, which according to Eq. (\ref{eqmulticontact}) is equal to 
\begin{equation}
I=-\frac{e^2}hT_2(V+V_s). 
\end{equation}
The impedance $Z_s$ between F1 and P is therefore 
\begin{equation}
\label{eqimps}Z_s=\frac VI=-\frac h{e^2}\frac 1{T_2}\frac{T_1+T_2}{2T_1+T_2}%
=-\frac h{2e^2}\left[ \frac 1{T_2}+\frac 1{2T_1+T_2}\right] . 
\end{equation}
For the (pedagogical) system with two fully polarized ferromagnets without spin-flip
processes the transmission possibility between the two ferromagnets is zero
when they are aligned antiparallel. The impedance $Z_s^{AP}$ is then simply
the reciprocal conductance between F1 and P 
\begin{equation}
Z_s^{AP}=-\frac h{e^2}\frac 1{T_2}. 
\end{equation}
The difference between the parallel and antiparallel configuration is 
\begin{equation}
\label{eqimpdifnf}Z_s^P-Z_s^{AP}=-Z_s^{AP}\frac{T_1^P}{2T_1^P+T_2}, 
\end{equation}
where $T_1^P$ is the transmission probability to F2 in the parallel
configuration.

In the presence of spin-flip scattering the transmission
probability $T_1^{AP}$ no longer vanishes. The difference in impedance between the parallel
and antiparallel configuration now becomes 
\begin{equation}
\label{eqimpdif}Z_s^P-Z_s^{AP}=\frac h{e^2}\frac{T_1^P-T_1^{AP}}{%
4T_1^PT_1^{AP}+2T_2(T_1^P+T_1^{AP})+T_2^2}. 
\end{equation}
Johnson \cite{citjohn}
implicitly assumed that the current into the drain contact did not affect
the impedance difference. This can be modelled by an infinite resistance to
the drain or $T_2=0$: 
\begin{equation}
\label{eqimpdifinfdrain}Z_s^P-Z_s^{AP}=\frac h{4e^2}\left[ \frac
1{T_1^{AP}}-\frac 1{T_1^P}\right] . 
\end{equation}
Eq. (\ref{eqimpdifinfdrain}) is equivalent to Johnson's result, which diverges in
the absence of spin-flip scattering
$T_1^{AP}\rightarrow 0$. 

Instead of the ill-defined $V_s$, Johnson measured a voltage $\widetilde{V}_s$ with
respect to a normal counter electrode $N$, {\em c.f.}
Fig.~\ref{spininj}(b). The transmission probabilities from 
$F1$ and $F2$ to the normal counterelectrode $N$ are called $T_{F1N}$ and $%
T_{F2N}$, respectively and $T_{NP}$ is the transmission probability from $N$
to $P$. Again setting the net current into F2 and N to zero:
\begin{equation}
\label{eqimp4}Z_s=\frac{\widetilde{V}_s}I=\frac h{e^2}\frac
1D(T_1T_{NP}-T_{F1N}T_2), 
\end{equation}
where 
\begin{equation}
\begin{array}{ccl}
D & = & T_2^2T_{NP}+T_2(T_2+T_{NP})(T_{F1N}+T_{F2N})+ \\  
&  & 2T_2T_{NP}T_1+(2T_2+T_{NP})(T_1T_{F1N}+T_1T_{F2N}+T_{F1N}T_{F2N}). 
\end{array}
\end{equation}
Assuming that the resistance to the
drain contact is relatively large, we can consider two limits.
When the normal counterelectrode N is positioned far from the drain
contact P both transmission probabilities to the drain contact, $T_2$ and $%
T_{NP}$, vanish and the impedance of the system reduces to 
\begin{equation}
\label{eqimpfarP}Z_s=\frac{\widetilde{V}_s}I=\frac h{3e^2}\frac{T_1-T_{F1N}}
{T_1(T_{F1N}+T_{F2N})+T_{F1N}T_{F2N}} .
\end{equation}
The impedance difference is 
\begin{eqnarray}
\label{eqimpdiff4} Z_s^P-Z_s^{AP}  =  \frac h{3e^2}
\frac{
(T_1^P-T_1^{AP})T_{F1N}(T_{F1N}+2T_{F2N})}{\left\{
T_1^P(T_{F1N}+T_{F2N})+T_{F1N}T_{F2N}\right\} } \nonumber \\
\times \frac{  1 } {\left\{
T_1^{AP}(T_{F1N}+T_{F2N})+T_{F1N}T_{F2N}\right\} } .
\end{eqnarray}
Comparing Eq. (\ref{eqimpdiff4}) with Eq. (\ref{eqimpdifinfdrain}), we
expect the four terminal configuration to be equivalent to the
three terminal configuration when $T_{F2N}\ll T_{F1N}$. However, the
impedance difference has a maximum for $T_{F2N}=0$: 
\begin{equation}
Z_s^P-Z_s^{AP}=\frac h{3e^2}\left[ \frac 1{T_1^{AP}} - \frac 1{T_1^P} \right] . 
\end{equation}
which is increased by a factor $4/3$ compared to the pedagogical model.

When the normal counterelectrode N is positioned close to the drain
contact P, $T_{NP}$ is no longer negligible and $T_{F1N}$ and $T_{F2N}$ are
small. The impedance difference 
\begin{equation}
Z_s^P-Z_s^{AP}=\frac h{e^2}\left[ \frac 1{T_1^{AP}} - \frac 1{T_1^P} \right] 
\frac{T_{F1N}T_{F2N}}{(T_{F1N}+T_{F2N})^2} 
\end{equation}
now depends on the ratio of $T_{F1N}$ and $T_{F2N}$. It has a maximum for $%
T_{F1N}=T_{F2N}$ for which it reduces to Eq. (\ref{eqimpdifinfdrain}).
Apparently, the impedance difference is affected by the choice of
position of the normal counterelectrode $N$.

A microscopic calculation of the various transmission probabilities involves
bulk, spin-dependent interface and spin-flip scattering processes. We present here a first
attempt based on a semiclassical calculation of a weakly scattering slice of a disordered
material.

\section{Sheet impurity scattering with spin-flip}

\label{scat}

In the effective mass approximation the
single-electron states at the Fermi energy $E_F$ are described by the
Schr\"odinger equation in a normal metal  
\begin{equation}
\label{eqschrodin}
\left[ \left( - 
\frac{\hbar ^2} {2m^{*}} \nabla^2 +V (\vec r) -E_F \right )  {\bf I} 
+ {\bf H}^{sg}(\vec r) + {\bf H}^{so}(\vec r) \right] 
\left( 
\begin{array} {c}
\psi_\uparrow (\vec r) \\
\psi_\downarrow (\vec r) 
\end{array} 
\right)
=0,
\end{equation}
where $m^{*}$ is the effective mass. We consider here
the effect of a thin sheet of short-range impurities with a scalar scattering
potential $V(\vec r)=\sum_\alpha {\bf \gamma }_\alpha \delta (z)\delta
(\vec \rho -\vec \rho _\alpha )$,
where $\vec \rho _\alpha $ gives the transverse position of the scattering
center, and ${\bf \gamma }_\alpha $ gives the strength of the scatterer.
Spin-flip scattering can be induced by
spin-orbit scatterers \cite{schiff}
\begin{equation}
\label{soschiff}
{\bf H}^{so}=\frac \hbar {4m^2c^2}\stackrel{%
\rightarrow }{\bf \sigma } \cdot 
\left[ 
\left( \stackrel{\rightarrow }{\nabla }V(\vec r) \right) 
\times 
\left( -i\hbar \stackrel{\rightarrow}{\nabla } \right)  \right] , 
\end{equation}
where $\overrightarrow{\sigma }$ is the vector containing the Pauli spin matrices, or
interaction with magnetic impurities with fixed random spin direction (spin glass):
\begin{equation}
\label{eq:Hpi}
{\bf H}^{sg} =\sum_{\alpha } J_{ex}\stackrel{\rightarrow 
}{S}_\alpha \cdot \stackrel{\rightarrow }{\sigma }\delta (z)\delta (\vec
\rho -\vec \rho _\alpha ). 
\end{equation}
Here $\stackrel{\rightarrow }{S}_\alpha $ is the spin of the paramagnetic
impurity and $J_{ex}$ is the local exchange integral. 
 
We wish to compute the Green function \cite{fisher,citbrata,citgijs}
\begin{equation}
G^{\pm }(\vec r,\sigma ;\vec {r}^{\prime },\sigma ^{\prime })=\sum_{\vec
{k_{\parallel }},\vec {k_{\parallel }^{\prime }}}G_{\vec k_{\parallel
}\sigma ,\vec k_{\parallel }^{\prime }\sigma ^{^{\prime }}}^{+}e^{ik_{\perp
}z_R}e^{i\vec k_{\parallel }\vec \rho _R}e^{-i{k^{\prime }}_{\perp
}z_L}e^{-i\vec k_{\parallel }^{\prime }\vec \rho _L}, 
\end{equation}
where $\vec r_{R,L}=( \vec \rho _{R,L}, z_{R,L})$ and $z_R>0$, $z_L<0$ are located in the
right and left leads. Without perturbations
\begin{equation}
\label{equnpgreen}
G_{\vec k_{\parallel
}\sigma ,\vec k_{\parallel }^{\prime }\sigma ^{^{\prime }}}^{+}
\rightarrow G_{\vec k_{\parallel }}^{+(0)} 
\delta_{\vec k _{\parallel },\vec k_{\parallel }^{\prime }}
\delta_{\sigma, \sigma ^{\prime }}
=-i\frac {m^*} {\hbar ^2}\frac
1{k_{\perp }}
\delta_{\vec k _{\parallel },\vec k_{\parallel }^{\prime }}
\delta_{\sigma, \sigma ^{\prime }}. 
\end{equation}
The matrix of transmission coefficients is related to the Green's function by
\cite{fisher} 
\begin{equation}
\label{eqtrgr}t_{\vec k_{\parallel }\sigma ,\vec k_{\parallel }^{\prime
}\sigma ^{^{\prime }}}=\frac{i\hbar ^2}m\sqrt{|k_{\perp }||{k^{\prime }}%
_{\perp }|}G_{\vec k_{\parallel }\sigma ,\vec k_{\parallel }^{\prime }\sigma
^{^{\prime }}}^{+}.
\end{equation} 
For isolated interfaces the following optical theorem holds \cite{citbrata}:
\begin{equation}
\label{eqGopt} \sum_{\sigma} \sum_{\vec k_{\parallel }}^{|\vec
k_{\parallel }|\le k_F} \left| G_{\vec
k_{\parallel }\sigma ,\vec k_{\parallel }^{\prime }\sigma ^{^{\prime
}}}^{+}\right| ^2 / G_{\vec k_{\parallel }}^{+(0)} =-i\mbox{Im}(G_{\vec
k_{\parallel }^{^{\prime }}\sigma ^{^{\prime }},\vec k_{\parallel }^{\prime}\sigma
^{^{\prime }}}^{+}). 
\end{equation}
We are interested in the transport properties averaged over impurity configurations. The
averaged Green function reads
\begin{equation}
\label{eqtrgr}
\langle G_{\vec k_{\parallel
}\sigma ,\vec k_{\parallel }^{\prime }\sigma ^{^{\prime }}}^{+}
 \rangle = \left( G_{\vec k _{\parallel }}^{+(0) \, -1} - 
\Sigma_{\vec k_{\parallel }} \right)^{-1} 
\delta_{\vec k _{\parallel },\vec k_{\parallel }^{\prime }}
\delta_{\sigma, \sigma ^{\prime }} . 
\end{equation}

The conductance can be calculated in two ways. By using Eq.
(\ref{eqGopt}) and the relation (\ref{eqtrgr}) between the Green's function and the
transmission coefficients: 
\begin{equation}
\label{eqcond}
G=\frac{2e^2}h   \sum_{\vec k_{\parallel }}^{|\vec k_{\parallel }|\le
k_{F}}\frac{ 1-(G_{\vec k_{\parallel }}^{+(0)})i\mbox{Im}\Sigma _{\vec k_{\parallel }}
}{|1-G_{\vec k_{\parallel }}^{+(0)}\Sigma _{\vec k_{\parallel }} |^2}, 
\end{equation}
Alternatively, the conductance can be calculated diagrammatically via the transmission
probabilities, which can be written in general as
\begin{eqnarray}
\label{eqtransprob}
\langle |t_{\vec k_{\parallel }\sigma ,\vec k_{\parallel
}^{\prime }\sigma ^{^{\prime }}}|^2\rangle & = & \frac 1{|1-G_{\vec k_{\parallel
}}^{+(0)}\Sigma _{\vec k_{\parallel }} |^2}\delta _{\vec k_{\parallel },\vec k_{\parallel
}^{^{\prime }}}\delta _{\sigma ,\sigma ^{^{\prime }}}
\\ \nonumber
& +& \frac{G_{\vec
k_{\parallel }}^{+(0)*}}{|1-G_{\vec k_{\parallel }}^{+(0)}\Sigma _{\vec k_{\parallel }} |^2}%
W_{\vec k_{\parallel }\sigma ,\vec k_{\parallel }^{\prime }\sigma ^{^{\prime
}}}\frac{G_{\vec k_{\parallel }^{\prime }}^{+(0)}}{|1-G_{\vec k_{\parallel
}^{\prime }}^{+(0)}\Sigma _{\vec k_{\parallel }} |^2}. 
\end{eqnarray}
Diffuse scattering and spin diffusion are described by the second term of Eq.
(\ref{eqtransprob}) which is governed by the (reducible) vertex correction $W$. 

These formally exact relations can be evaluated for the Born approximation which is valid
for low impurity densities
$n_{IR}=N_{IR}/A$ and weak scattering strength. With an average strength of the scatterers  $\overline{\gamma } =
\sum_{\alpha } \gamma _{\alpha ,\sigma }/N_{IR}$ and a mean square value $\gamma  ^2= \sum_{\alpha }
\gamma _{\alpha ,\sigma }^2/N_{IR}$, the self-energy reads in the Born approximation
 \begin{equation}
\label{eqselfB}\Sigma ^B=n_{IR}\overline{\gamma } -i\frac{\hbar ^2}
{m^*} k_F(\eta +\eta _{sf}), 
\end{equation}
where $\eta$ is the spin-conserving scattering parameter 
\begin{equation}
\label{eqscatpar}
\eta =\frac{n_{IR}}{2\pi }\left( \frac{
m^* \gamma  }{\hbar ^2}\right) ^2. 
\end{equation}
$\eta _{sf}=\eta _{sf}^{so}+\eta _{sf}^{sg}$ is the spin-flip scattering parameter caused by
spin-orbit scatterers with
$u_{so}=
\hbar ^2 k_F^2 / 4 m^2 c^2$ 
\begin{equation}
\eta _{sf}^{so}=\frac 23\frac{n_{IR}}{2\pi }\left( \frac{m^* u_{so}\gamma }{
\hbar ^2}\right) ^2 
\end{equation}
and by paramagnetic impurities 
\begin{equation}
\eta _{sf}^{sg}=\frac{n_{IR}}{2\pi }\left( \frac{m^* J}{\hbar ^2}\right) ^2. 
\end{equation}
Assuming $ \rm{Re}\Sigma ^B \ll \rm{Im}
\Sigma ^B$ and inserting Eq. (\ref{equnpgreen})
into Eq. (\ref{eqcond}) 
\begin{equation}
G=\frac{2e^2}h    \sum_{\vec k_{\parallel }}^{|\vec
k_{\parallel }|\le k_F}\frac{k_{\perp }}{k_{\perp
}+(\eta +\eta _{sf})k_F}. 
\end{equation}
To the first order in $\eta $ the conductance is given by 
$  G/{G_0}= 
1-2(\eta +\eta _{sf})$ , 
where 
$G_0=(2e^2/h)(Ak_F^2/4\pi ) $
is the Sharvin conductance. In order to calculate the spin diffusion, we
need the transmission probability for spin-flip scattering. To the lowest
order in the scattering parameters
the transmission probability matrix in spin space is obtained by summing $\langle
|t_{\vec k_{\parallel },\vec k_{\parallel }^{\prime }}|^2\rangle $ over the incoming and
outgoing states. The spin-flip probability $T_{sf}$ under transmission is
\begin{equation}
\label{eqTsf}
T_{sf}=\frac{Ak_F^2}{2\pi }\frac 23\eta _{sf}. 
\end{equation}
whereas the spin-conserving transmission probability $T $ is 
\begin{equation}
\label{eqTnf}T =\frac{Ak_F^2}{2\pi }\left( \frac 12-(\eta +\frac 53\eta _{sf})\right) . 
\end{equation}

\section{Finite thickness}

\subsection{Relevant length scales}
\label{lengths}

Next to the geometrical parameters of a sample, {\em i.e.} the sample cross sections $A$
and length $L$, several characteristic length scales govern the transport properties.
In the (quasi-) ballistic regime the conductance is dominated by
the contact resistance.  In the diffuse regime
transport is limited by scattering at bulk impurities. An obvious parameter is the impurity scattering {\it mean free path}
$\ell$.
The {\it spin-flip mean free path }$\ell _{sf}$ 
is the average
length an electron travels before it flips its spin. 
Including scattering at spin-orbit scatterers or paramagnetic impurities,
does not only decrease $\ell _{sf}=v_{F }\tau _{sf }$, 
where $\tau _{sf}$ is the spin-flip scattering time and $v_F$ the Fermi velocity,
but also reduces the
mean free path ${\ell  }^{-1}=v_{F
}^{-1} (\tau ^{-1} +\tau _{sf}^{-1}) $.
In the diffuse regime the relevant parameter is not $\ell _{sf}$, but the {\it spin
diffusion length } $l_{sf}=\sqrt{\ell \ell
_{sf}/6}$ over
which the spin-accumulation persists \cite{citvf}.

\subsection{Two-terminal conductance}
\label{bulkmaterial}

The transport properties of samples with  finite thickness can be readily obtained from the
results of the previous chapter in two different limits. When the mean-free path is much
larger than the current path, the system is in the quasiballistic limit and the results
above can be carried over directly. In the
quasi-ballistic regime, $L\ll \ell $, we can make the connection of the scattering
mean free paths and the microscopic parameters 
$N(\eta+\eta_{sf})=L/2 \ell$ and
$N \eta_{sf}=L/2 \ell_{sf}$.
The conductance is
$G^{qb}(L)/G_0 =1-2L/\ell$. The spin-flip
probability for transmission through a thin slice of bulk material follows from Eq.
(\ref{eqTsf}):
\begin{equation}
\label{eqTsfN}T_{sf}^{qb}(L)=\frac{Ak_F^2}{2\pi }\frac 23N\eta _{sf}=\frac{Ak_F^2}{%
4\pi }\frac L{\ell _{sf}} 
\end{equation}
and for transmission without spin-flip 
\begin{equation}
\label{eqTnfN}T^{qb} (L)=\frac{Ak_F^2}{4\pi }\left( 1-\frac L{\ell 
}-\frac L{\ell _{sf}}\right) . 
\end{equation}
In the diffuse regime, where $L\gg \ell $ or $N(\eta +\eta
_{sf})\gg 1$, the conductance is 
$G^{df}(L)/G_0= 4 \ell/3L$.
The spin-resolved two-terminal transmission probabilities in the diffuse
regime, on the other hand, can be obtained from the scattering properties of a thin slice by
following Schep {\it et al.}
\cite{citschep}. It is shown in \cite{linda} that this procedure is equivalent to solving the
diffusion equation \cite{citvf,citfertlee} with the following general solutions for the
spin-average $\overline{\mu }=Az+B$ and the spin-splitting $\delta \mu$ of the chemical
potentials:
\begin{equation}
\label{solunequi}\delta \mu (z)=C\exp (- z/l_{sf})+D\exp (z/l_{sf}) .
\end{equation}
The integration constants $A,B,C,D$ have to be determined by
the boundary conditions. The transport properties are easily obtained from the chemical
potentials. 

\section{Spin transistor}

We can now collect the different results to obtain expressions for the
spin transistor. We approximate the transmission probabilities in the many terminal
configuration by those in the two-terminal configuration. This is not unproblematic because
we do not know the current paths. Complication are for example
spreading resistances at contacts, which reflect the fact that electrons do not always
travel straight between the two contacts. However, in a thin film geometry like the spin
transistor, we expect the corrections to be not too large for a qualitative study. It should
be kept in mind that the geometric parameters, {\em i.e.} the cross section $A$ and the
current path length $L$, may somewhat deviate from the geometrical measures of the sample. 

\subsection{Quasi-ballistic regime}

Let us consider first the quasiballistic regime and assume that the contacts are made from
strong ferromagnets in which the minority carrier density vanishes and the majority
density equals that of the normal metal island. In that limit the performance of the
spin-transistor is determined by the spin-flip scattering length $\ell _{sf} > \ell$.
From Eqs. (\ref{eqTsfN}) and (\ref{eqTnfN}), the transmission probabilities
are 
\begin{equation}
\label{qb1}
T_1^P=\frac{Ak_F^2}{4\pi }\left( 1-\frac{L_1}\ell -\frac{L_1}{\ell _{sf}}%
\right) 
\end{equation}
and 
\begin{equation}
\label{qb2}
T_1^{AP}=\frac{Ak_F^2}{4\pi }\frac{L_1}{\ell _{sf}}. 
\end{equation}
The transmission probability to the drain contact can be either
quasi-ballistic or Ohmic, depending on the distance $L_2$ between the
ferromagnetic and drain contact. If the drain channel is quasi-ballistic ($%
L_2\ll \ell $) 
\begin{equation}
T_2=\frac{Ak_F^2}{4\pi }\left( 1-\frac{L_2}\ell \right).
\end{equation}
The impedance difference for the pedagogical model is to lowest order in $L_1$:
\begin{equation}
\label{eqzball}\left[ Z_s^P-Z_s^{AP}\right] ^{qb}=\frac 1{G_0}\frac{3L_2}{%
8\ell }\left[ 1-\frac{3L_2}{4\ell }\frac{L_1}{\ell _{sf}}\right] .
\end{equation}
This equation also holds when transport to the drain is diffusive.

\subsection{Diffuse regime}

Let us now consider tunnel contacts with a spin-injection 
efficiency which can be expressed in terms of $G_\sigma ^F$, the tunneling
conductance for spin $\sigma $, respectively the density of states at the Fermi energy
$N_{\sigma }(E_F)$: 
\begin{equation}
\label{eqeffpolg}\beta =\frac{N_{\uparrow }(E_F)-N_{\downarrow }(E_F)}{%
N_{\uparrow }(E_F)+N_{\downarrow }(E_F)}=\frac{G_{\uparrow }^F-
G_{\downarrow
}^F}{G_{\uparrow }^F+G_{\downarrow }^F}. 
\end{equation}
We will now consider a system consisting of a normal metal of thickness $L_N$ sandwiched
between two ferromagnetic tunneling junctions, over which a voltage
$\Delta \mu /e$ is applied.  Defining the average tunneling conductivity 
$G_F=2G_{\uparrow }^F G_{\downarrow }^F  /( G_{\uparrow }^F+G_{\downarrow}^F )$ 
we can solve the diffusion equations, with the result
for the transmission coefficients: 
\begin{equation}
\label{eqTP4}\frac 1{T_1^P}=\frac{L_1}{4g_N}+\frac 12\frac{\sinh (L_N/2l_{sf})+(1-\beta
^2)\frac{g_N}{l_{sf}}\frac 1{G_F}\cosh (L_N/2l_{sf})}{G_F\sinh
(L_N/2l_{sf})+ g_N \cosh (L_N/2l_{sf})/l_{sf}} 
\end{equation}
\begin{equation}
\frac 1{T_1^{AP}}=\frac{L_1}{4g_N}+\frac 12\frac{\cosh (L_N/2l_{sf} )+(1-\beta ^2)\frac 1{G_F}\frac{g_N}{l_{sf}}\sinh
(L_N/2l_{sf})}{G_F\cosh (L_N/2l_{sf})+ g_N \sinh (L_N/2l_{sf})/l_{sf}} 
\end{equation}
\begin{equation}
\label{eqTFN4}\frac 1{T_{FN}}=\frac{L_2}{4g_N}+\frac 14
\frac
{\sinh ( L_2/l_{sf})+(1-\beta ^2)\frac 1{G_F} \frac {g_N}{l_{sf}} \cosh (L_2/l_{sf})}
{G_F\sinh (L_2/l_{sf})+ g_N \cosh (L_2/l_{sf})/l_{sf}} 
\end{equation}
where $g_N =Ak_F^2 \ell/ 3 \pi$. 
In the limit $T_2 \rightarrow 0$ 
the impedance difference becomes 
\begin{equation}
\label{impdiffpart}Z_s^P-Z_s^{AP}=\frac{2\rho _N}A\frac{\beta ^2l_{sf}}{\left[ \left( 
l_{sf}G_F/g_N \right) ^2+1\right] \sinh (L_1/l_{sf} 
)+2G_F l_{sf} \cosh (L_1/l_{sf})/g_N} 
\end{equation}
has a maximum for $G_F=0$: 
\begin{equation}
\label{max}
Z_s^P-Z_s^{AP}=\frac{2\rho _N}A\frac{\beta ^2l_{sf}}{\sinh (L_1/
l_{sf})}, 
\end{equation}
which agrees with \cite{citfertlee} in the tunneling limit. 

By an analysis of his experiments Johnson found $l_{sf}=1.5\pm
0.4\mu$m \cite{citjohnscience}. The impedance of the spin transistor is approximately 25 $\mu \Omega $ in
the parallel configuration and the measured impedance is 3 $\mu \Omega $ for
the 1.6 $\mu $m thick film. The largest impedance difference predicted for
the optimum value $\left| \beta \right| =1$, using Eq. (\ref{max})
and the above value found for $l_{sf}$, is $Z_sAd\approx 3\times 10^{-2}$ $\Omega $
$\mu $m$^3$, which is about three times {\it smaller} than the measured
difference. This problem might be caused by contamination which reduces the contact area.

\section{Conclusions}

We analyzed the spin transistor by a semiclassical scattering theory of transport resolving some
inconsistencies which arise from a two-terminal approximation for a many-terminal device.
The performance of the spin transistor is limited by the spin-diffusion length in the diffuse regime and the
spin-flip scattering length in the quasi-ballistic regime. Naturally, we many expect a better transistor
action in clean samples.

\section*{Acknowledgments}

This work is part of the research program for the ``Stichting voor
Fundamenteel Onderzoek der Materie'' (FOM), which is financially
supported by the ''Nederlandse Organisatie voor Wetenschappelijk
Onderzoek'' (NWO).  This study was supported by the NEDO joint
research program (NTDP-98).  We acknowledge benefits from the TMR
Research Network on ``Interface Magnetism'' under contract No.
FMRX-CT96-0089 (DG12-MIHT).  We also acknowledge stimulating
discussions with K.M. Schep, J.~Caro, P.~J.~Kelly and Yu.V. Nazarov.

%
%

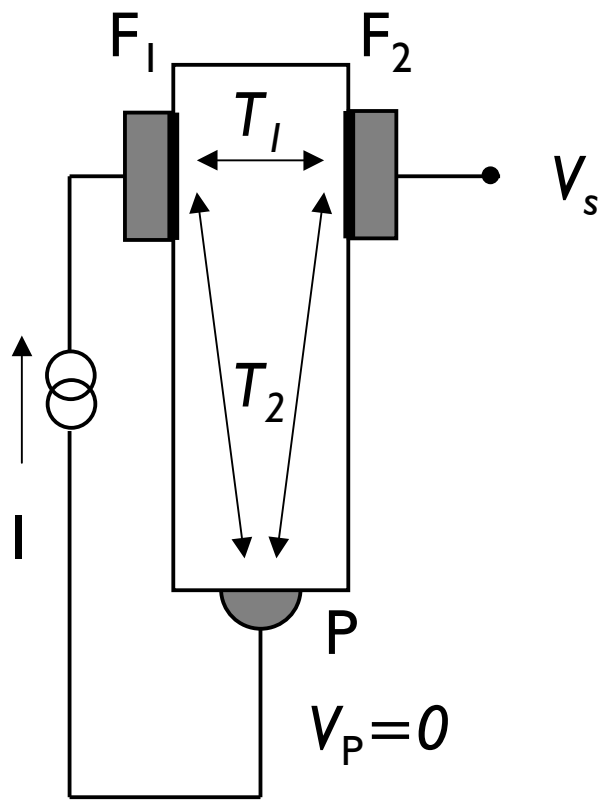
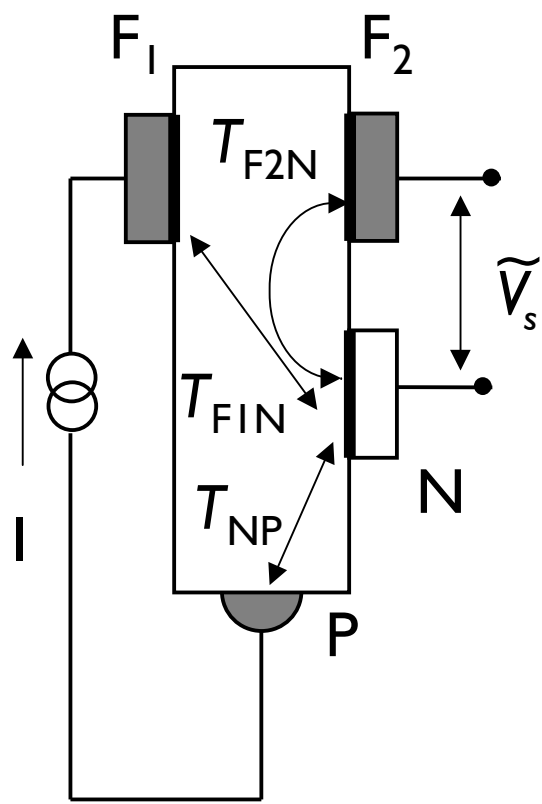
\end{document}